\documentclass[aps,prl,10pt,twocolumn,superscriptaddress]{revtex4-1}
\usepackage{graphicx}
\usepackage{amsmath}
\usepackage{amssymb}

\begin{document}
\title{Large Drag Reduction over Superhydrophobic Riblets}
\author{Charlotte Barbier}\thanks{Address all correspondence to this author.}
    \affiliation{
	Computational Sciences and Engineering Division,
	Oak Ridge National Laboratory,
	Oak Ridge, Tennessee 37831}

\author{Elliot Jenner}
\author {Brian D'Urso}
	\affiliation{Department of Physics \& Astronomy,
	University of Pittsburgh,
	Pittsburgh, Pennsylvania, 15213
    }
\begin{abstract}

Riblets and superhydrophobic surfaces are two demonstrated passive drag reduction techniques. We describe a method to fabricate surfaces that combine both of these techniques in order to increase drag reduction properties. Samples have been tested with a cone-and-plate rheometer system, and have demonstrated significant drag reduction even in the transitional-turbulent regime. Direct Numerical Simulations have been performed in order to estimate the equivalent slip length at higher rotational speed. The sample with 100~$\mu$m deep grooves has been performing very well, showing drag reduction varying from 15 to 20~$\%$ over the whole range of flow conditions tested, and its slip length was estimated to be over 100~$\mu$m.

\end{abstract}
%
%

\maketitle

\section*{INTRODUCTION}

Techniques for reducing hydrodynamic drag are of great interest since they can lead to improved energy efficiency and performance for a wide range of applications. Passive methods such as permeable walls \cite{Hahn02}, or riblets \cite{Walsh90} can reduce the hydrodynamic drag on a flat plate by up to 10~$\%$. Even better results have been obtained with active methods such as polymer injection \cite{Virk75}, bubbles injection \cite{Sanders06}, and air layers \cite{Fukuda00,Elbing08} which are capable of reducing frictional drag by as much as 80\%. However, none of these active methods have been successfully transferred to engineering applications: the polymers' efficiency degrades at high strain rate, and injecting micro-bubbles or sustaining an air layer on a non-flat surface is too challenging. With the progress in nanotechnology, a new passive method arose in the last decade: superhydrophobic surfaces \cite{Ou04,Ou05,Balasubramanian04,Gogte05,Lee08,Lee09}. The reader is referred to Rothstein \cite{Rothstein10} for a recent comprehensive review on the research on drag reduction on superhydrophobic surfaces. On these surfaces, the contact angle of a water droplet generally exceeds $150^\circ$ and the roll off angle is less than $10^\circ$. Thus, they are extremely hard to wet and are highly hydrophobic. Typically, they consist of a combination of small structures with dimensions and spacing ranging from $100$~nm to $10$~$\mu$m and a hydrophobic material or coating. Abundant examples can be found in nature such as plant leaves (lotus, tulip tree, or eucalyptus), insect wings and legs (water strider), and bird feathers.

\begin{figure}[ht]
\centering
	\includegraphics[width=3in]{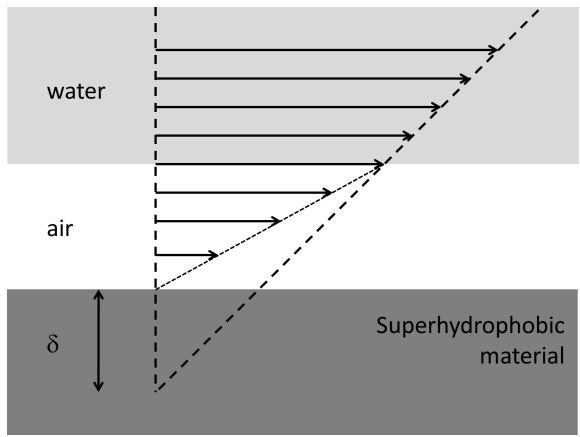}
	\caption{Schematic of the effective slip on a superhydrophobic surface. The velocity at the water-air interface is defined as the slip velocity.\label{fig:slip_def}}
\end{figure}

The drag reduction mechanism of superhydrophobic surfaces is similar that of the air layer method. A thin film of gas is held on the surface, and the large viscosity difference between the air and water can be translated into slip boundary condition at the wall characterized by a slip length $\delta$ (see Fig. \ref{fig:slip_def}). The slip velocity $U_s$ can be then defined with the slip length and the shear rate experienced at the wall such as:
\begin{equation}
U_s = \delta \frac{dU}{dn}
\label{eq:slipBC}
\end{equation}
It is generally accepted that the larger the slip length, the lower the skin friction. Superhydrophobic surfaces are capable of reducing drag over a wide range of flow \cite{Ou04,Ou05,Balasubramanian04,Gogte05,Choi06a,Choi06b} and their limitation lies in their capacity to hold the air layer at high shear rate. In this paper, we combined riblets similar to the work of Choi et al. \cite{Choi93}, streamwise grooves, with superhydrophobic surfaces in order to achieve large slip effects. The main goal was to design a surface that would demonstrate large drag reduction in turbulent flow. A spike geometry was chosen rather than typical microposts in order to minimize the liquid-solid area as suggested by \cite{Ybert07}. We also conjecture that this geometry makes it more difficult for the flow to dislodge the bubbles trapped in it. 

. The grooves, or riblets, contribute to the drag reduction in several ways: (a) they align the turbulent vortices near the surface and limit the vortex interaction; (b) they limit the drag increase caused by slip effects in the spanwise direction observed in several studies~\cite{Min04,Fukagata06}; and (c) a larger air layer may be trapped in the bottom of the grooves. The drag reduction properties of the surfaces are measured with a cone-and-plate rheometer over a wide range of rotational speed. The fabrication process of the samples and the experimental testing method are fully described in this paper. In order to estimate the slip length at higher rotational speed, Direct Numerical Simulations (DNS) of the cone-and-plate flow were performed for three slip lengths and several rotational speeds.

\section*{MATERIALS AND METHODS}
\subsection*{Fabrication of superhydrophobic samples}

\begin{figure}
	\centering
	\includegraphics[width=3in]{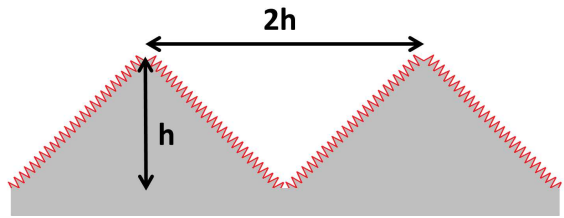}
	\caption{Cross section schematic of one sample (not at scale).\label{fig:grooves_schematic}}
\end{figure}

\begin{figure}
	\centering
	\includegraphics[width=3in]{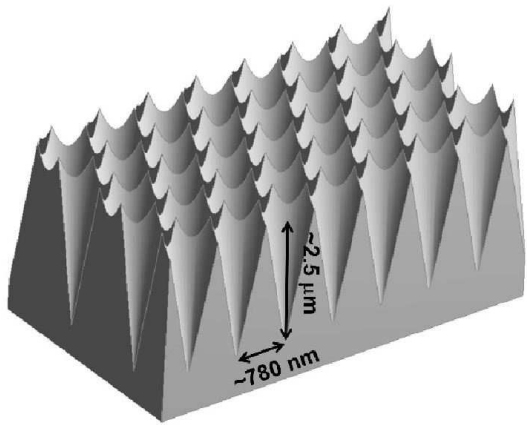}
	\caption{Schematic and dimensions of the flared pores.\label{fig:pores_schematic}}
\end{figure}

\begin{figure}
	\centering
	\includegraphics[width=3in]{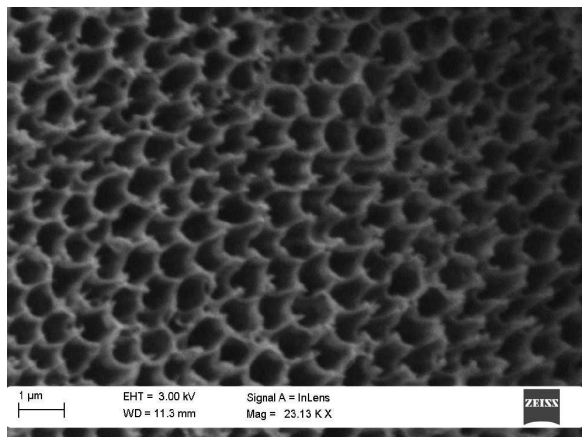}
	\caption{Scanning Electron Microscope (SEM) images of the superhydrophobic surfaces made by repeated anodization and etching of aluminum with $10$~$\mu$m grooves.\label{fig:SEM_10um}}
\end{figure}

The fabrication of our superhydrophobic materials was similar to the multi-step anodization and etching process used by Masuda~\cite{Masuda07} to make molds for nanoimprint lithography of tapered surface features. A 99.99\% purity 4 inch diameter aluminum disk (Lesker EJTALXX402A4) was annealed and cut flat ($\sim 10$~nm roughness) by single point diamond turning. Next, concentric grooves were cut into the aluminum using a 90 degree dead sharp diamond tool, forming a series of sharp ridges and grooves. Three different depths $h$ (see Fig. \ref{fig:grooves_schematic}) of grooves were tested: $0$~$\mu$m (flat surface), $10$~$\mu$m (flat sample), $100$~$\mu$m, and $1,000$~$\mu$m. The micro-structure which forms the superhydrophobic surface was created by a series of anodizing steps in citric acid, alternated with etching in tetramethylammonium hydroxide. The anodizing steps create aluminum oxide pores with $\sim 780$~nm spacing (determined by  anodization voltage). These pores grow into the aluminum substrate, while the etching widens the pores at each step. The combined effect results in flared aluminum oxide pores, where the pores are wide at the surface, and narrow as they go deeper into the substrate. As the anodizing-etching steps continue, the pores start intersecting between themselves and leaves spikes in a hexgagonal arrangement, as shown in Fig. \ref{fig:pores_schematic} and Fig. \ref{fig:SEM_10um}. More detailed explanation on the fabrication process can be found in \cite{Jenner13}. To render the surfaces superhydrophobic, the samples were spin-coated with a sub-$\mu$m thick layer of Hyflon AD60, a hydrophobic polymer, which conformally coated the structure. The samples are thus composed of small spikes with $\sim 780$~nm spacing located on grooves that are from one to three order of magnitude larger (see Fig. \ref{fig:grooves_schematic}).

\begin{figure}
\centering
	\includegraphics[width=3in]{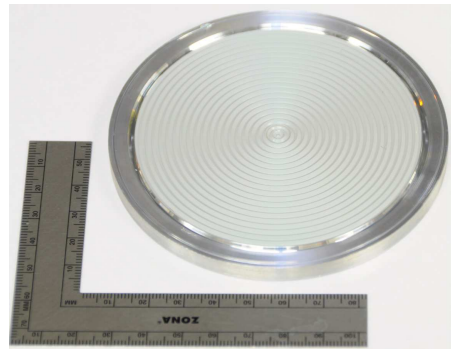}
\caption{Photograph of a multiscale superhydrophobic surface with 1~mm deep grooves.\label{fig:1mm_grooves}}
\end{figure}

\begin{figure}
\centering
	\includegraphics[width=3in]{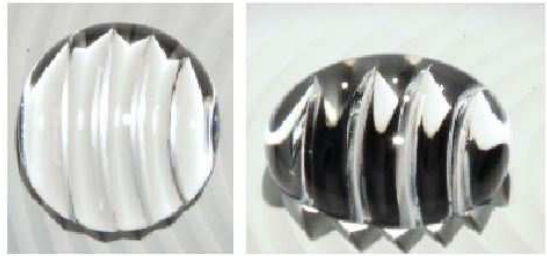}
\caption{Top and side picture of a large drop on the 1~mm deep grooves sample\label{fig:drop_pic}}
\end{figure}

The nano structures needed for the superhydrophobic surface can be generated on any aluminum substrate, whether it is flat, grooved, or any other complex shape, which is the key advantage of this fabrication method. Furthermore, the nano pores are always perpendicular to the substrate surface as a consequence of the anodization process, guaranteeing a high quality superhydrophobic surface. The combination of nano pores and the Hyflon coatings was found to be quite robust \cite{Jenner13} and is an excellent choice for drag reduction techniques. A photograph of the sample with $1$~mm grooves is shown in Figs.~\ref{fig:1mm_grooves} and \ref{fig:drop_pic}. The superhydrophobic properties of the samples were checked on the flat sample: a static angle of $161^\circ \pm 1$ and an average rolling angle of $0.2^\circ$ for $15 \mu  L$ drops \cite{Jenner13}.

\subsection*{Experimental approach}
\begin{figure}
	\centering
	\includegraphics[width=3in]{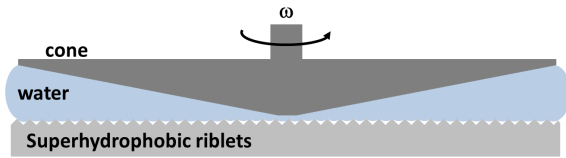}
	\caption{Schematic of the experimental setup.
	\label{fig:setup}}
\end{figure}

The samples and a control disk (no grooves and no hydrophobic coating) were tested with a commercial cone-and-plate rheometer (AR 2000, TA Instruments) at the Center for Nanophase Materials Science (CNMS) at Oak Ridge, TN, USA. The rheometer can measure torque as low as $10^{-7}$~N.m with a resolution of $10^{-9}$~N.m. The samples were used as the bottom plate, whereas a  stainless-steel cone with a 60 mm diameter, $2^\circ$ angle, and 51~$\mu m$ in truncation was used to generate the flow motion, as shown in Fig.~\ref{fig:setup}. Distilled water was pipetted with an exact volume of $1.98 \pm 0.01$~mL on to the superhydrophobic sample. Then, the cone was slowly lowered until contact, and then was raised 51~$\mu m$. Due to the truncation, the cone tip diameter is approximately 2.9~mm wide, which is large enough to land on the top of the grooves for all the samples tested here, ensuring a good axial position of the cone.  For the 100 and 1,000 ~$\mu$m deep grooves samples, some water was ejected from the gap between the cone and the sample. The excess of water was removed with a cotton swab, paying attention that the meniscus remained in a good shape. The final volume of water between the cone and the sample was estimated to be between 1.95 and 1.99~mL. Since the torque is porportional to the volume of water in laminar regime, the error on the water volume can lead to an error of 2~$\%$ on the torque. For higher rotational speeds, the error on the torque can be estimated with the empiral formula given in \cite{Sdougos84}, and it was found to be less than 3.2~$\%$;this was the main source of error. A first series of measurements was performed at low rotational speed, from 2 to 6~rad/s, with a 0.5~rad/s increment, in order to test the drag reduction properties in the laminar regime. A second set of measurements at higher rotational speed was performed starting from 6~rad/s, with a 4~rad/s increment, continuing until the water was squeezed out of the cone-and-plate region (from 54 to 70 rad/s depending on the samples). Viscous heating during measurements can affect the water viscosity, and thus reduce the torque on the cone. For instance, a $0.1^\circ$C temperature increase causes an underestimation of the torque by 1~$\%$. However, the Brinkmann number, $Br = \mu r^2 \omega^2/kT$, which defines the tendency of the system to show viscous heating effects, is very small: $2.4 \times 10^{-5}$. Using the work of Turian and Bird \cite{Turian63}, we estimated that the viscous heating could cause an error of less than 0.0005~$\%$, and thus was negligible. Although slight drag reduction were observed with the flat sample (ranging from 0 to 4 $\%$), the measurements are not reported since most of them were within the margin of errors of the torque measurements.

\subsection*{Numerical approach}

The flow in a cone-and-plate has been extensively investigated due to its application in viscosity measurements. In rheology, the cone-and-plate is generally used with a low rotational speed and a small cone angle resulting in a laminar flow. However, the cone-and-plate apparatus has also been shown to be well suited for studying well-characterized turbulent flow fields \cite{Einav94}. Cone-and-plate flow is characterized by a single parameter $\tilde{R}=\frac{r^2 \omega \alpha^2}{12 \nu}$ \cite{Sdougos84}, where $r$ is the radial distance, $\omega$ the rotational speed, $\alpha$ the cone angle, and $\nu$ the kinematic viscosity. This is the ratio of the centrifugal force to the viscous force acting on the fluid. When $\tilde{R}$ is small enough, the centrifugal force is negligible and the radial component of the flow is zero everywhere, causing the streamlines to be concentric circles. This flow is often referred as the primary flow. As $\tilde{R}$ increases, a secondary flow appears: a radial fluid motion appears due to the no-longer negligible centrifugal force, causing the streamlines to cease being concentric. Experimental and theoretical models are available to estimate this secondary flow \cite{Sdougos84}. As $\tilde{R}$ becomes larger than $\sim 4$ , turbulence appears at the outer edge of the disk. For water and our cone geometry, $\tilde{R}=4$ occurs at about $44$~rad/s.

At low rotational speed, when no secondary flow is present, the torque is:

\begin{equation}
\begin{split}
T = \frac{2\pi \omega \mu R_0^3}{3\alpha} \left( \vphantom{\frac{3\delta^2}{R_0^2\alpha^2}} 1 \right.&-\frac{3\delta}{2R_0\alpha}+\frac{3\delta^2}{R_0^2\alpha^2} \\
& \left. \vphantom{\frac{3\delta^2}{R_0^2\alpha^2}}-\frac{3\delta^3}{R_0^3\alpha^3}ln\left(1+\frac{R_0\alpha}{\delta}\right) \right) \\
\end{split}
\label{eq:torqueLaminar}
\end{equation}

where $\mu$ is the dynamic viscosity, and $R_0$ is the cone radius.

However, Eq.~(\ref{eq:torqueLaminar}) is no longer valid when secondary flow appears. In order to estimate the slip length at higher rotational speed, Direct Numerical Simulations (DNS) were performed with OpenFOAM (Open Source Field Operation and Manipulation) \cite{OpenFoam}, an open-source computational Fluid Dynamics (CFD) program. The transient incompressible Navier-Stokes equations were solved using the PISO algorithm \cite{Ferziger02}. The geometry consists of the region between the cone and the plate from $r$ = 1.46 to 30~mm, the lower limit corresponding to the cone truncation. The flow between r = 0 and 1.46~mm is not modeled for several reasons: (a) the problem of the singularity at r = 0 is removed; (b) it avoids the presence of very small cells; (c) the torque caused by the shear on this small region is much smaller (only 0.01~\%) than over the rest of the region; and (d) the flow in this region remains laminar over the whole range of rotational speed investigated. To guarantee high quality simulation, the grid spacing was chosen based on Kolmogorov's length scale following Yeung and Pope's suggestion \cite{Yeung89}. For instance, for the simulations at 80~rads/s, the mesh was composed of 105 million cells. The simulations were run on Frost \footnote{http://www.nccs.gov/computing-resources/frost/} at ORNL, an SGI Altix ICE 8200 cluster node with quad-core Intel-Xeon processors and 24 GB memory. The following boundary conditions were used: (a)~circumferential velocities at the cone and at the radius r = 1.46 mm taken as $r\omega$; (b)~a shear-free condition for the free surface at the outer rim; and (c)~a slip condition with a constant slip length on the flat surface. Rotational speeds of 2, 5, 6, 10, 20, 40, 60 and 80~rad/s were used and three slip lengths were tested: 0 (i.e. no slip), 100, and 200~$\mu$m. Note that the main objective of the simulation is to give a rough estimate of the slip length in the higher rotational speeds, when the flow is no more a simple primary flow. There is no evidence that the slip length is uniform accross the sample and that the slip length boundary condition used in the simulation is adequate at higher rotational speed. Ffor instance, the air-water interface may not remain flat and non-negligible air flow may occur in teh grooves. However, since the slip length is usually used by material scientists to describe the slip property of superhydrophobic surface, the DNS results are used only for this purpose and no other data such as turbulence information have been extracted from the results.   

\section*{RESULTS}

\subsection*{Results at low rotational speed (from 2 to 6~rad/s)}
\begin{figure}
\centering
	\includegraphics{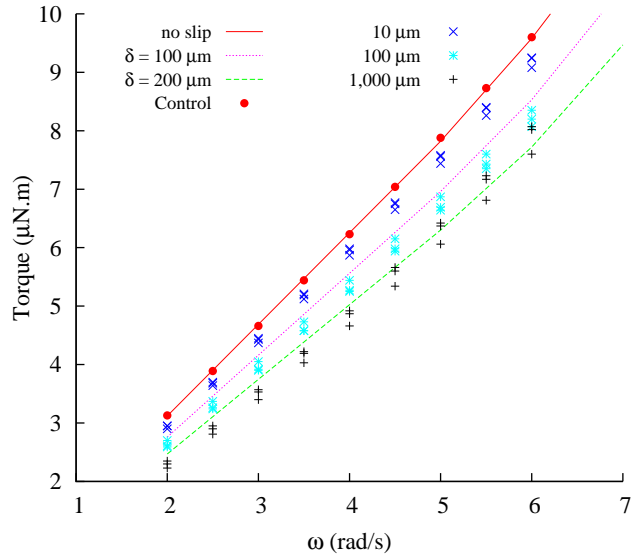}
\caption{Torque measured on the cone at low rotational speed for different samples (markers) and torque computed with the CFD simulations for different slip lengths.\label{fig:laminarTorque}}
\end{figure}

In this range of rotational speed, the flow is laminar, but secondary flow appears at about 4~rad/s. Fig. \ref{fig:laminarTorque} shows the torque applied on the cone for each sample, and the corresponding results. Lower torque is observed for samples with larger grooves, which can be explained by the presence of a thicker air layer. It is generally accepted that surfaces with larger slip length will cause larger slip in the laminar regime \cite{Rothstein10}. As can be seen in Fig. \ref{fig:drop_pic}, the water is laying on the top of the grooves, and the grooves remain dry. As a consequence, the slip length is larger for samples with larger grooves, resulting in greater drag reduction.

\begin{figure}
\centering
	\includegraphics{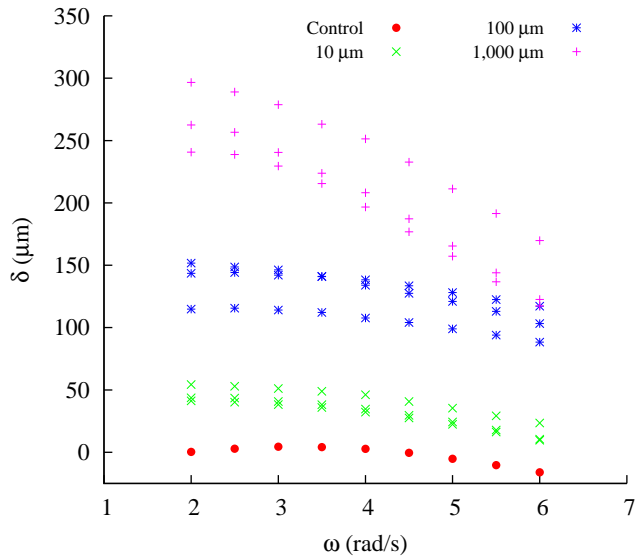}
\caption{Slip lengths calculated with Eq.~(\ref{eq:torqueLaminar}) in the laminar regime for the control disk and the samples with 10, 100, and 1,000 $\mu$m grooves.\label{fig:laminarDelta}}
\end{figure}

Fig. \ref{fig:laminarTorque} shows good agreement between the simulations and the experiment (less than 0.7~$\%$ deviation), and suggests that the slip length for 1,000~$\mu$m deep grooves is about 200~$\mu$m. The slip length can also be estimated using expression (\ref{eq:torqueLaminar}), and the corresponding slip lengths are plotted in Fig. \ref{fig:laminarDelta}. Note that Eq.~(\ref{eq:torqueLaminar}) is not valid in the presence of secondary flow, which will cause an increase in the torque. This is observed on the control sample for rotational speeds above 4 rad/s where the slip length becomes negative. Good agreement is found between the slip length from Eq.~(\ref{eq:torqueLaminar}) and the DNS simulations shown in Fig. \ref{fig:laminarTorque} up to 4~rad/s; for instance, the slip length with 1,000~$\mu$m deep grooves is found to be larger than 200~$\mu$m, and the slip length with 100~$\mu$m deep grooves ranges from 100 to 200~$\mu$m.

\begin{figure}
\centering
	\includegraphics{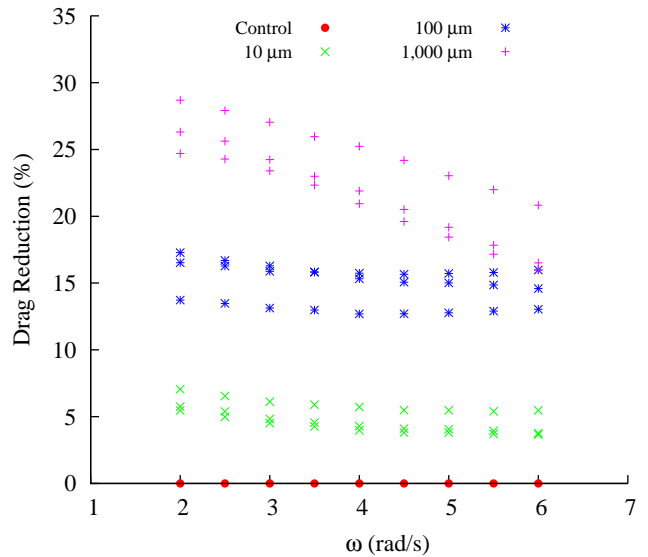}
\caption{Drag reduction (in $\%$) compared to the control flat sample in laminar regime.}
\label{fig:DRlaminar}
\end{figure}

The drag reduction properties of each sample are estimated by comparing the torque on the cone with the control sample, $DR(\%) =100 (T_{control}-T_{sample})/T_{control}$, shown in Fig. \ref{fig:DRlaminar}. The largest drag reduction was observed with 1,000~$\mu$m deep grooves, up to 29\% at 2 rad/s, but the efficiency decreases with increasing rotational speed. Performance of the 10~$\mu$m and 100~$\mu$m deep grooved samples is more constant over this range of rotational speeds, and drag reduction of approximately 4~\% and 15~\% was observed, respectively. The higher drag reduction with 1,000~$\mu$m deep grooves can be exlpained by the large quantity of air trapped inside the grooves (see Fig. \ref{fig:drop_pic}). The deviation in the measurements comes mainly from excess water removed for larger grooves sample when the cone was slowly lowered.

\subsection*{Results at high rotational speed (from 6 to 70~rad/s)}
\begin{figure}[t]
\centering
	\includegraphics[width=3in]{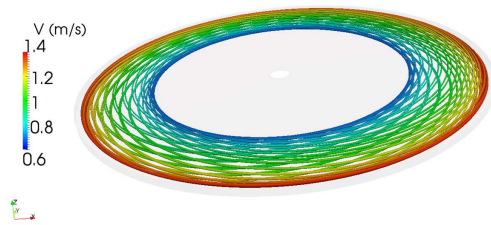}
\caption{Streamlines obtained with the simulations at 60~rad/s with a 200~$\mu$m slip length with the instantaneous flow fields.}
\label{fig:3Dstream}
\end{figure}

\begin{figure}[t]
\centering
	\includegraphics{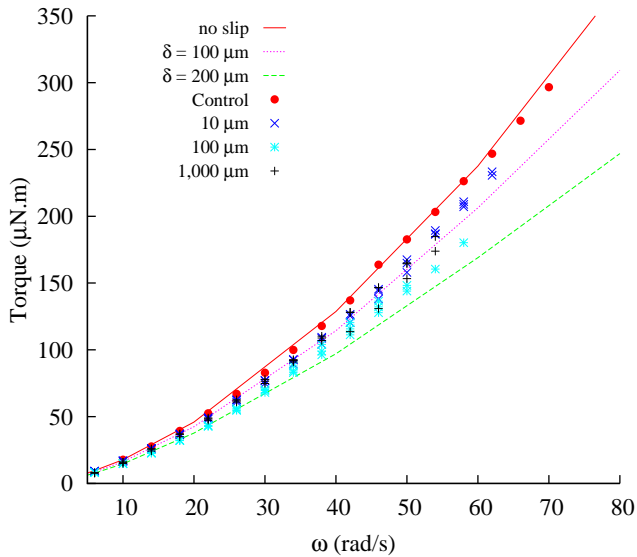}
\caption{Torque measured on the cone at low rotational speed for different samples (markers) and torque computed with the CFD simulations for different slip lengths}
\label{fig:turbulentTorque}
\end{figure}

In this range of rotational speed, the flow is no longer laminar and the transition to the turbulent regime occurs at 44 rad/s at the outer edge. As the rotational speed increases, the torque on the cone ceases to be linear with respect to the rotational speed: secondary flow and turbulence result in larger drag (see Fig. \ref{fig:turbulentTorque}). The secondary flow is well illustrated in Fig. \ref{fig:3Dstream}, which shows the three dimensional streamlines obtained with the simulations at 60~rad/s with a 200~$\mu$m slip length.

Similar to the low speed regime, very good agreement is observed between the CFD simulations and the control sample, with less than 1 \% difference. The measurements stopped when the water was flung out of the cone-and-plate space due to the centrifugal force: $\sim 70$~rad/s for the control sample, $\sim 62$~rad/s for the 10~$\mu$m deep groove sample, $\sim 58$~rad/s for the 100~$\mu$m deep grooves sample, and $\approx 54$~rad/s for the 1,000~$\mu$m deep groove sample. The water may have been flung out more easily from the more deeply grooved sample because of the presence of a thicker air layer, which enables easier radial motion. At 44~rad/s, the torque on the cone for the control disk seems to be abnormally large, however this may be caused by the transition to turbulence. From Fig. \ref{fig:turbulentTorque}, the slip lengths of the 10 and 1,000~$\mu$m deep groove samples can be roughly estimated to be between 50 and 100~$\mu$m, while the slip length of the 100~$\mu$m deep groove sample is $\approx$ 150~$\mu$m.

\begin{figure}[t]
\centering
	\includegraphics{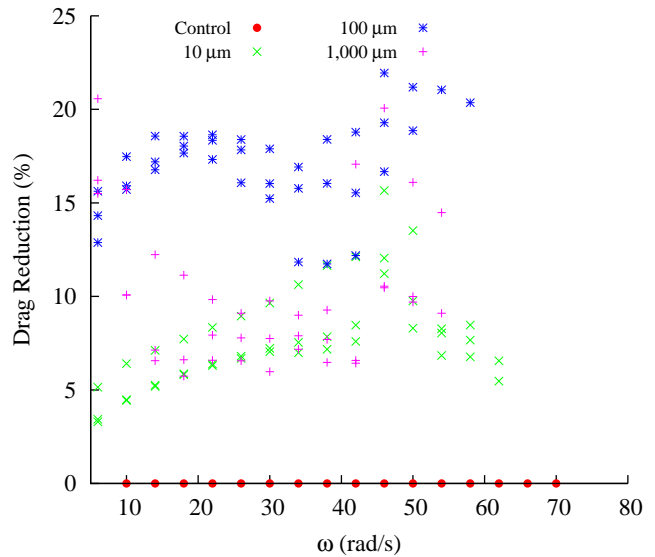}
\caption{Drag reduction (in~$\%$) compared to the control flat sample in higher rotational speed.}
\label{fig:DRturbu}
\end{figure}

Similarly to the low speed regime, the drag reduction properties of each sample relative to the control disk are plotted in Fig. \ref{fig:DRturbu}. The best results were observed for the 100~$\mu$m deep groove sample, which reduced the drag by up to 20~\%, even at high rotational speed. The 10 and 1,000~$\mu$m deep groove samples were capable of reducing the drag by 5~\% to 10~\% only. Note that larger drag reduction was observed for all the samples at 44~rad/s, which corresponds to the transition to turbulence for the cone-and-plate flow \cite{Sdougos84}. This larger drag reduction may be caused by a delay in the transition to turbulence, which would result in lower torque measurement. 

The results of Choi et al. \cite{Choi93} shows that overly large grooves could induce a drag increase, whereas smaller grooves could reduce drag by aligning the streamwise vortices above the surface.  Although our grooves are not filled with water amd straight like in \cite{Choi93}, similar behavior is observed with our samples, which suggests a common drag reduction mechanism. More precisely, for the 100~$\mu$m grooved sample, the shape factor  $s^+$ \cite{Choi93} of groove spacing over the boundary layer thickness at 60~rad/s is approximately 8, which is small enough to cause drag reduction \cite{Choi93}. 

\section*{CONCLUSIONS}
An innovative surface was designed to efficiently and passively reduce the drag over a large range of flow regimes. The combination of grooves and a superhydrophobic surface was capable of reducing the drag up to 20~$\%$ in turbulent regime. The experiments showed that if the grooves were too small or too large, the drag reduction would be reduced but still present (at least 5~$\%$).

As an example of application, using the formula of Fukagata et al. \cite{Fukagata06}, a 300~m oil tanker cruising at 16 knots would have its drag reduced by at least 44~\%  if it had a hull made of such material. However, the slip length of our samples were measured with a shear rate of up to 1,700~s$^-1$ which is still one order of magnitude lower than in a tanker flow ($\approx 5\times 10^4 s^{-1}$).

The results above prompt further investigation of this material in the fully turbulent flow regime, and development of better experimental methods to investigate drag reduction properties at higher flow regimes. 

\begin{acknowledgments}

A portion of this research was conducted at the Center for Nanophase Materials Sciences, which is sponsored at Oak Ridge National Laboratory by the Scientific User Facilities Division, Office of Basic Energy Sciences, U.S. Department of Energy. This research was supported by ORNL Seed Money Program. This manuscript has been authored by employees of UT Battelle, LLC, under contract DE-AC05-00OR22725 with the U.S. Department of Energy. Accordingly, the United States Government retains and the publisher, by accepting the article for publication, acknowledges that the United States Government retains a non-exclusive, paid-up, irrevocable, world-wide license to publish or reproduce the published form of this manuscript, or allow others to do so, for United States Government purposes.
\end{acknowledgments}

\bibliography{biblio}

\end{document}